# E-learning for ungraded schools of Kazakhstan: experience, implementation, and innovation

*Kerimbayev Nurassyl, Akramova Aliya, Suleimenova Jarkynbike*

*Kazakhstan,Astana*

**Abstract**

The modernization of the educational process in the ungraded schools of the Republic of Kazakhstan requires the provision of affordable quality education for students in rural areas on information technology, the creation of e-learning. It was important to consider two points: how does e-learning influence the educational process in ungraded schools, and directly on the quality of teaching, and what results can thus be achieved. The significance of the work is also to explore innovative approaches to e-learning system of ungraded schools.

**The key issue**

Each student in any country has the constitutional right to a quality education. We asked the question: Are these rights implemented for students of rural areas, whether they get the same amount of knowledge like in innovative schools in the big city, whether there is the same access to advanced information technology in education? Having based on the differences in the level of knowledge, skills and competences in general, we have found that children have different levels of knowledge when they finish the secondary school and unequal starting level for admission to higher education. This led us to the need to develop a set of measures to promote the implementation of quality education for students in ungraded schools of the countryside.

### 1.Introduction

One of the key objectives of Education System Informatization of the Republic of Kazakhstan is creation of a unified educational information environment, e-learning, which provide students with equal opportunities for acquiring knowledge at all levels of education. One of the priorities of the educational policy of the Republic of Kazakhstan is the development of ungraded schools. Today ungraded schools represent 70% of the total number of schools in Kazakhstan. Migration to the cities, the demographic situation contribute to the preservation of such schools, and more than that, an increase in their number.

We give the following definition for ungraded schools. Ungraded schools - is secondary school with a small contingent of students, a combined class-sets and a specific form of training sessions.

The idea of creating an electronic system for ungraded schools is a promising direction in the development and improvement of education systems in many countries.

The works studying the problems of rural schools, the creation of conditions for their development and improvement are analyzed in research journals. Zammit (Zammit, K. (2011) analyzes the experience of teachers in rural areas on the border of New South Wales and Victoria. Connecting multi literacies and engagement of students from low socio-economic backgrounds, using Bernstein's pedagogic discourse as a bridge, (Language and Education, 25 (3), 203-220.) Kennedy and Cavanaugh conducted the study on a virtual school. (Kathryn Kennedy, Cathy,.2013).

In the Republic of Kazakhstan, as well as in many other countries, the government pays great attention to the problems of ungraded schools, their development and modernization.

The purpose of our work is to investigate the data for the implementation and organization of e-learning system in ungraded schools of the Republic of Kazakhstan. It was necessary to consider what kind of results and outcomes were achieved.

### II. Methodology

To determine the basis of e-learning, it is necessary to consider the theoretical and methodological aspects of the issue, and to develop a methodology for self-study. The methodological tools were a theoretical analysis of the literature survey and interviews, the analysis of the results, forecasting the future direction of research.

When conducting an interview with the director of the ungraded school (village Taskudyk, South Kazakhstan) Mr. B.Taubayev, he was asked the following questions: what are the problems experienced in ungraded schools for today? Mr D. Taubayev identified the following challenges facing the ungraded schools:

1. A focused training of staff for ungraded schools. The fact is that many teachers do not know or do not come across with the specific work terms for ungraded school in their activities.

2. Available modern computer equipment, necessary technical equipment of ungraded schools and unpreparedness of teachers to use it.

3. "Closed" work of ungraded schools due to its distance from other population centers and localities, lack of integration with other educational, scientific and public organizations.

Today, all of these problems and conflicts can be solved with the help of information and communication and networking technologies. We see a solution to this problem in the implementation of e-learning system for ungraded schools.

### III. Main part
### III.1 What is e-learning for ungraded schools?

The primary objectives facing us in the development of scientific-methodological bases for the use of e-learning and the creation of an integrated information system for educational ungraded schools were:

analysis of characteristics of the models of ungraded schools, its structure, functions (teaching, diagnostic, compensatory, adaptive, cognitive, etc.) and training principles, taking into account the usage of computer technology;

• characteristics of information and communication technology and educational tools, their ability to provide training in ungraded schools, peculiarities of teacher activities on the application of information and communication technology;

• creation and implementation of e-learning systems taking into account characteristics of the ungraded school.

The main target guideline of e-learning for ungraded schools was the creation of conditions aimed at:

• improving the quality of education through the effective use of modern information and communication technologies;

• improving the quality of education by providing access of remote rural schools to educational resources;

• creation of e-learning materials, methodical software.

To implement the e-learning system in ungraded schools successfully, internal readiness and ability to perceive the achievements of scientific and technical progress and effective use of these advances for educational and training challenges facing the school are necessary.

The creation and implementation of the models of e-learning systems in a holistic educational process of ungraded primary school should improve the competitiveness and contribute to its development.

The development of this model implies:

– determining the current state of ungraded schools, to define the term "ungraded schools";

– determining scientific-methodological basis of the learning process taking into account the characteristics of the ungraded schools (formation of class-sets, various options of class-sets, one-disciplinary and multi-disciplinary lessons in different types of class-sets);

– application of information and communication technology, network resources to distance education, and online resources;
– creation of digital educational resources (adapted electronic textbooks for students in ungraded schools, electronic workbooks, test systems, virtual labs, etc.);
– training teachers of ungraded schools to apply computer technology in teaching and learning.

Reforming and optimization of ungraded schools requires the introduction of new technologies, tools and forms of education in the educational process, as well as readiness of teachers to apply them in their work.

Creation and implementation of e-learning contributes to the development and education of the intellectual potential of individuals. Electronic system for ungraded schools should be characterized by accessibility, openness, and focused on Web-technology and software products. Other features are complexity, security technical support, flexibility to adapt.

Creation of an e-learning system involves the development of scientific and methodological bases of information and computer technologies use in ungraded schools, its structures, functions and principles of learning in the educational institutions of this type.

Kazakhstan has the experience in the establishment and operation of ungraded schools, which implement the idea of a unified educational environment, the system of e-learning. A model of the "School of Information" has been created and implemented which operates in eastern Kazakhstan in Stepnovsk school from 2005. In this model, the school widely uses e-learning technology (in the form of keys and network) saturated with electronic teaching resources (electronic multimedia textbooks and manuals, videos) that allows teachers to manage the activities of the combined class-sets, in particular, self-activities of people with the assistance of learning computer programs.

The use of case and the network technology is optimal for learning in ungraded schools. There is an interactive relationship between teachers and students with the network technology. Here the communication can be built on-line, and in external form. Student chooses a task that is located on the site (portal), performs it, and sends it to the teacher. The marks are recorded in the electronic journal, access to which is open to both students and parents. Online communication can be carried out with the help of skype, audio, using text information in the form of chat.

E-learning system in ungraded schools solves the problems of informatization of education, acts as part of the educational infrastructure of the information space of the Republic of Kazakhstan.

**III.II e-learning system for small schools of the Republic of Kazakhstan: the technology of implementation.**

Electronic system in ungraded schools has been implemented by us by informatization of education, which include computer-hardware, software, and their substantive content.

Currently, there are mainline data links in the republic, providing high-speed exchange of large flows of information. The possibilities of such information and communication environment are used at Abay Kazakh National Pedagogical University, which is implementing the "Virtual School Academy" project. The prerequisite for video conferencing with ungraded schools of Kazakhstan is the availability of the communication channel and the capacity of the respective bands.

The e-learning system in ungraded schools of the Republic of Kazakhstan is carried out within the framework of this project. We conducted on-line lessons, Internet-conferences and round tables with students in ungraded schools.

Using the module Conference Moderator provided performing video conferencing sessions via the Web-based interface with ungraded schools.

The scheme of video conferencing with ungraded schools of the Republic of Kazakhstan is as follows:

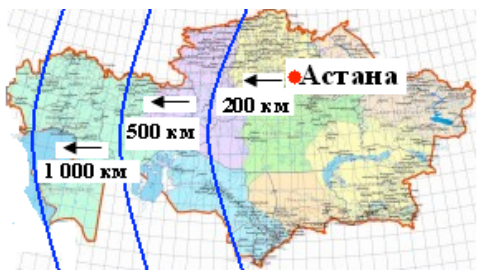

Fig. 1. Organization of video conferencing with ungraded schools of the Republic of Kazakhstan

Video conferencing system includes the hardware and software components. According to the number of participants videoconferences were as follows:
- Personal (dialogue between the two parties);
- Group (communication between groups of participants);
- Studio (one speaker to the audience.)

Videoconference organized by Open Meetings - server for the local network or the Internet, which allows to use an Internet browser, the Adobe Flash Player plugin. Open Meetings – server for the local network or the Internet - used for conferences.

Open Meetings allows you to use the Internet browser plug-in Adobe Flash Player, creating different types of audiences. Members of videoconferencing can display documents, use graphics, control your TV screen and video conferencing. (Kerimbayev, 2013)
E-learning in ungraded schools can be presented as follows:

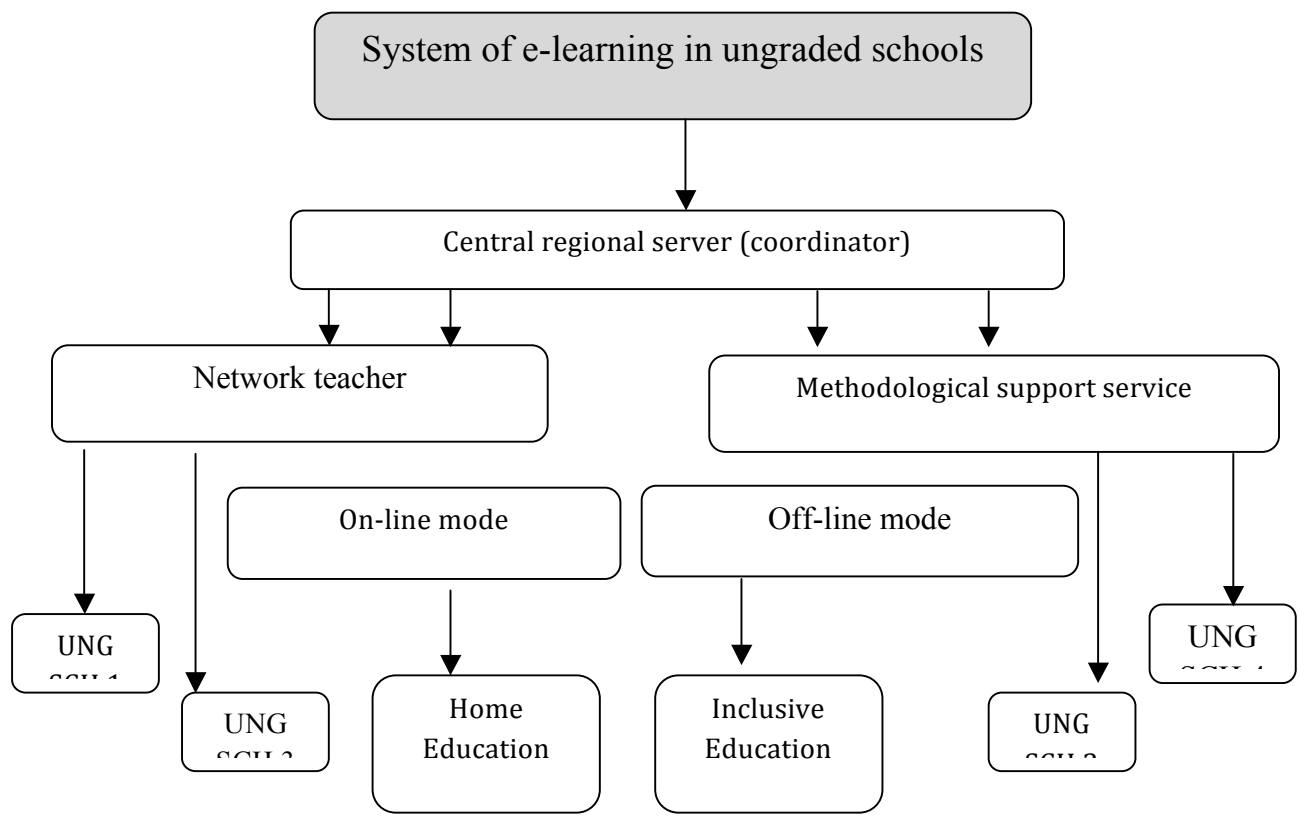

Fig. 2 The e-learning in ungraded schools

The e-learning system in ungraded schools can provide interactive learning process, structure and specific content of its modularity, the repetition of the material studied several times, self-control and analysis of educational achievements, the privacy of an individual learning process.

Distance learning, conducted in remote access, requires the organization of high-quality sound and picture. This can be achieved by using high quality cameras and audio communication devices that provide Wi-Fi-quality audio and full-motion video.

Thus, one of the ways of ungraded schools is the creation and implementation of e-learning system. The electronic learning (e-Learning) is able to reduce the significant shortcomings of the educational process in ungraded schools.

In the development of the e-learning it is important to highlight the following aspects into account:
- Methodological;
- Economics;
- Technology;
- Technology;
- Methodical.

Video conferencing and webinars were directed to work with both students and the teachers to improve their qualifications, experience sharing, etc. Organized "round tables", the meeting with the Methodists, scientists, authors of textbooks and teaching aids, of course, expanded horizons of teachers, increased their professionalism and professional competence. Conducting lessons online with students of ungraded schools increased their educational interest and activity. Information and communication technology training requires skills of the teachers to apply computer technology and the availability of a certain amount of skill of the computer students. Practice and experience show that even elementary school learners possess basic skills and abilities of the computer use, and sometimes even more aware of the latest news in modern technology than adults. Today there is enough experience on the computer technology training use.

The issue of the use of ICT (information-communication technologies) in education in ungraded primary school remains open and requires the exchange of experiences, views and pedagogical ideas in this area. However, the very necessity of the use of computer technology in education is undeniable, and proved that the computer technology learning enhances students' interest in learning, stimulates mental activity, and involves students in learning and cognitive activity.

Good technical equipment of schools, including the connection of a computer class to Internet channel facilitates ICT learning in primary school, opens the possibility of combining many subjects, allows students to fully realize their creative potential. Integrated training is fully implemented with application of computer technology in education that allows to move from an isolated consideration of the various phenomena of reality to their interconnected, complex study.

**V. Conclusion**

Today to improve the methods and technologies of the educational process in ungraded schools, the structure and model of learning in ungraded schools using computer technology (the creation of educational information resources and management of the educational process in the small schools in the network mode, etc.) are one of the topical issues.

A range of problems outlined by the director of ungraded school, Mr. Taubayev, can be solved by the introduction of e-learning systems. According to our study, e-learning solves a number of problems, including: improving the quality of education in ungraded schools, professional development of teachers, cooperation and collaboration of ungraded schools with other organizations.